\documentclass[runningheads]{llncs}
\usepackage{overpic}
\usepackage{amsmath}
\usepackage{amssymb}
\usepackage{graphicx}
\newcommand{\eg}[1]{\textit{e.g.,}}
\newcommand{\ie}[1]{\textit{i.e.,}}
\newcommand{\etc}[1]{\textit{etc}}
\usepackage{threeparttable}
\usepackage{color}
\usepackage{multirow}
\newcommand{\figref}[1]{Fig.\!~\ref{#1}}
\usepackage[table]{xcolor}
\usepackage{colortbl}
\definecolor{bblue}{rgb}{0,150,230}
\definecolor{mygray}{gray}{.9}
\definecolor{myy}{RGB}{126,95,0}
\usepackage{bm}

\usepackage[pagebackref=false,breaklinks=true,colorlinks=true,bookmarks=false,citecolor=blue]{hyperref}

\begin{document}
\title{Multi-Contrast MRI Super-Resolution via a Multi-Stage Integration Network}  
%
%
\author{Chun-Mei Feng\inst{1,2}
\and
Huazhu Fu\inst{2}
\and
Shuhao Yuan\inst{2}
\and
Yong Xu \inst{1}}
\institute{$^1$ Shenzhen Key Laboratory of Visual Object Detection and Recognition, Harbin Institute of Technology, shenzhen\\
$^2$ Inception Institute of Artificial Intelligence \\
{\tt \{strawberry.feng0304\}@gmail.com}\\
{\tt \href{https://github.com/chunmeifeng/MINet}{https://github.com/chunmeifeng/MINet}}
\thanks{This work was done during the internship of C.-M.~Feng at Inception Institute of Artificial Intelligence. Yong Xu is the corresponding author.}
}
%
\maketitle              
\begin{abstract}
Super-resolution (SR) plays a crucial role in improving the image quality of magnetic resonance imaging (MRI). MRI produces multi-contrast images and can provide a clear display of soft tissues. However, current super-resolution methods only employ a single contrast, or use a simple multi-contrast fusion mechanism, ignoring the rich relations among different contrasts, which are valuable for improving SR. In this work, we propose a multi-stage integration network (\ie, MINet) for multi-contrast MRI SR, which explicitly models the dependencies between multi-contrast images at different stages to guide image SR. In particular,  our MINet first learns a hierarchical feature representation from multiple convolutional stages for each of different-contrast image. Subsequently, we introduce a multi-stage integration module to mine the comprehensive relations between the representations of the multi-contrast images. Specifically, the module matches each representation with all other features, which are integrated in terms of their similarities to obtain an enriched representation. Extensive experiments on fastMRI and real-world clinical datasets demonstrate that 1) our MINet outperforms state-of-the-art multi-contrast SR methods in terms of various metrics and 2) our multi-stage integration module is able to excavate complex interactions among multi-contrast features at different stages, leading to improved target-image quality.


\end{abstract}
\section{Introduction}
Magnetic resonance imaging (MRI) is a popular technique in medical imaging. Unlike other modalities, such as computed tomography (CT) or nuclear imaging, MRI can provide clear information about tissue structure and function without inducing ionizing radiation. However, due to the complex data acquisition process, it is difficult to obtain high-resolution (HR) MRI images~\cite{feng2021brain,feng2021accelerated}.
%

Since the super-resolution (SR) can improve the image quality without changing the MRI hardware, this post-processing tool has been widely used to overcome the challenge of obtaining HR MRI scans~\cite{feng2021T2Net}. Bicubic and b-spline interpolation are two basic SR interpolation methods~\cite{dong2015image,feng2021brain}; however, they inevitably lead to blurred edges and blocking artifacts. Relying on the inherent redundancy of the transformation domain, iterative deblurring algorithms~\cite{hardie2007fast,tourbier2015efficient}, low rank~\cite{shi2015lrtv}, and dictionary learning methods~\cite{bhatia2014super} have made significant progress in MRI SR. Recently, deep learning which offers higher resolution, has also become widely used for the task~\cite{oktay2016multi,pham2017brain,mcdonagh2017context,chen2018brain,chaudhari2018super,feng2021DualOctConv,feng2021DONet}. For example, Akshay \textit{et al.}~applied a 3D residual network to generate thin-slice MR images of knees~\cite{chaudhari2018super}. Chen \textit{et al.}~used a densely connected SR network to restore HR details from a single low-resolution (LR) input image~\cite{chen2018brain}. Lyu \textit{et al.}~used a generative adversarial network (GAN) framework for CT denoising and then transferred it to MRI SR~\cite{lyu2018super}. However, the above methods focus on mono-contrast acquisition to restore HR images, ignoring complementary multi-contrast information. 


MRI produces multi-contrast images under different settings but with the same anatomical structure, \eg, T1 and T2 weighted images (T1WI and T2WI), as well as proton density and fat-suppressed proton density weighted images (PDWI and FS-PDWI), which can provide complementary information to each other. 
For example, T1WI describe morphological and structural information, while T2WI describe edema and inflammation. Further, PDWI provide information on structures such as articular cartilage, and have a high signal-to-noise ratio (SNR) for tissues with little difference in PDWI, while FS-PDWI can inhibit fat signals and highlight the contrast of tissue structures such as cartilage ligaments~\cite{chen2015accuracy}. In clinical settings, T1WI have shorter repetition time (TR) and echo time (TE) than T2WI, while PDWI usually take shorter than FS-PDWI in the scanning process. Therefore, the former are easier to acquire than the latter. These can be employed to add further information to a single LR image. For instance, relevant HR information from T1WI or PDWI can be used to assist the generation of SR T2WI or FS-PDWI~\cite{xiang2018deep}. Recently, several methods for multi-contrast SR have been proposed~\cite{zheng2017multi,zheng2018multi,zeng2018simultaneous,lyu2020multi}. For example, Zheng \textit{et al.}~used the local weight similarity and relation model of the gradient value between  images of different contrast to restore an SR image from its counterpart LR image \cite{zheng2017multi,zheng2018multi}. Zeng \textit{et al.}~proposed a deep convolutional neural network to simultaneously obtain single- and multi-contrast SR images \cite{zeng2018simultaneous}. Lyu \textit{et al.}~introduced a progressive network, which employs a composite loss function for multi-contrast SR \cite{lyu2020multi}. However, though significant progress has been made, existing methods do not explore the interaction between different modalities in various stages, which is especially critical for the target-contrast restoration.


%

To explore the correlations among hierarchical stages of different contrast, we propose a multi-stage integration network (MINet). In our method, the features of each stage are interacted and weighted. Specifically, the complementary features are fused with the corresponding target features, to obtain a comprehensive feature, which can guide the learning of the target features. Our main contributions are as follows: 1) We design a novel multi-contrast SR network to guide the SR restoration of the target contrast through the HR auxiliary contrast features. 2) We explore the response of multi-contrast fusion at different stages, obtain the dependency relationship between the fused features, and improve their representation ability. 3) We perform extensive experiments on fastMRI and clinical datasets, demonstrating that our method can obtain superior results compared with the current state of the arts.

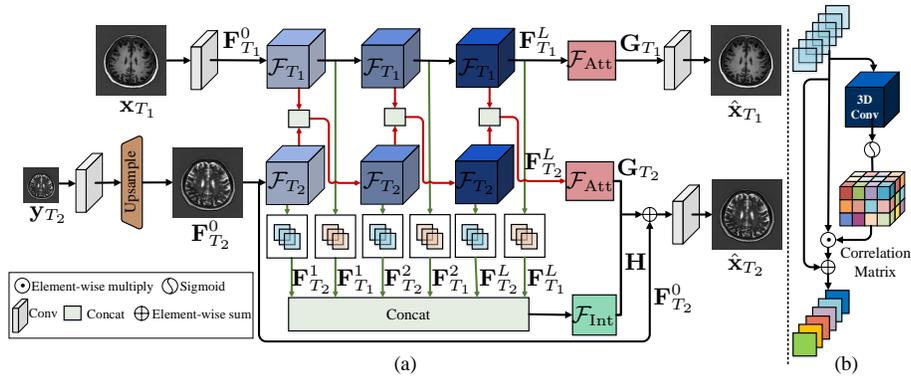
\begin{figure}[!t]
\centering
  \includegraphics[width=0.99\textwidth]{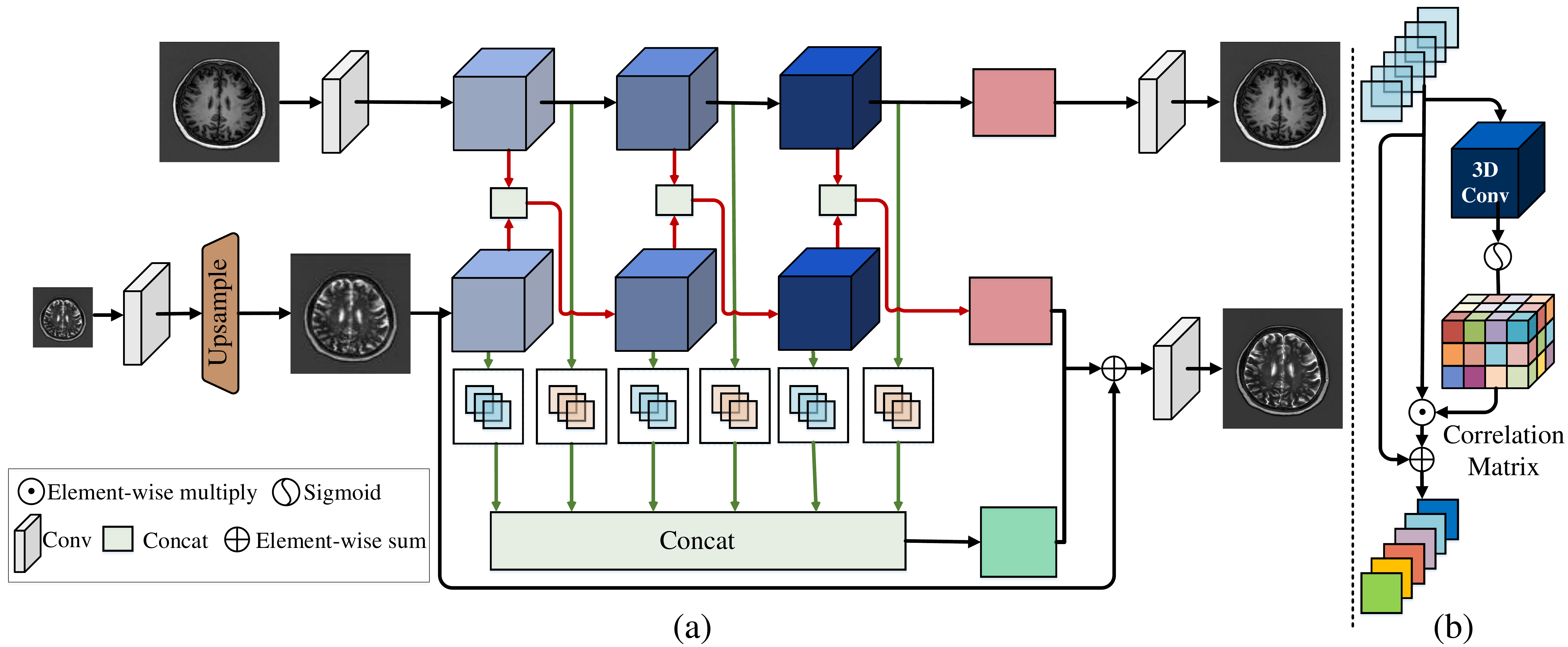}
  \put(-300,100){\footnotesize $\mathbf{x}_{T_1}$}
  \put(-334,60){\footnotesize $\mathbf{y}_{T_2}$}
  \put(-70,98){\footnotesize $\hat{{\mathbf x}}_{T_1}$}
  \put(-70,40){\footnotesize $\hat{{\mathbf x}}_{T_2}$} 
  \put(-244,114){\footnotesize $\mathcal{F}_{T_1}$} \put(-208,114){\footnotesize $\mathcal{F}_{T_1}$}
  \put(-172,114){\footnotesize $\mathcal{F}_{T_1}$}
  \put(-244,71){\footnotesize $\mathcal{F}_{T_2}$}
  \put(-208,71){\footnotesize $\mathcal{F}_{T_2}$}
  \put(-173,71){\footnotesize $\mathcal{F}_{T_2}$}
  \put(-131,71){\footnotesize $\mathcal{F}_{\text{Att}}$}
  \put(-131,117){\footnotesize $\mathcal{F}_{\text{Att}}$}
  \put(-129,20){\footnotesize $\mathcal{F}_{\text{Int}}$}
  \put(-112,76){\footnotesize $\mathbf{G}_{T_{2}}$}
  \put(-111,123){\footnotesize $\mathbf{G}_{T_{1}}$}
  \put(-109,40){\footnotesize $\mathbf{H}$}
  \put(-98,27){\footnotesize $\mathbf{F}_{T_{2}}^{0}$}
  \put(-235,34){\footnotesize $\mathbf{F}_{T_2}^1$}
  \put(-218,34){\footnotesize $\mathbf{F}_{T_1}^1$}
  \put(-200,34){\footnotesize $\mathbf{F}_{T_2}^2$}  
  \put(-182,34){\footnotesize $\mathbf{F}_{T_1}^2$}  
  \put(-164,34){\footnotesize $\mathbf{F}_{T_2}^L$}  
  \put(-146,34){\footnotesize $\mathbf{F}_{T_1}^L$}  
  \put(-261,124){\footnotesize $\mathbf{F}_{T_1}^0$}
  \put(-273,51){\footnotesize $\mathbf{F}_{T_2}^0$}
  \put(-147,78){\footnotesize $\mathbf{F}_{T_{2}}^L$}
  \put(-149,124){\footnotesize $\mathbf{F}_{T_{1}}^L$}
  \caption{(a) Network architecture of the proposed multi-contrast MRI SR model. (b) Details of the channel-spatial attention module (\ie, $\mathcal{F}_{\text{Att}}$).} 
  \label{figure1} 
\end{figure} 
\section{Methodology}

\subsection{Overall Architecture} 
Given an LR image $\mathbf{y}\in\mathbb{R}^{N\times N}$, we aim to learn a neural network that can provide an HR image $\mathbf{x}\in\mathbb{R}^{M\times M}$ ($M>N$). While mainstream efforts address this task by directly restoring $\mathbf{x}$ from $\mathbf{y}$, we propose to solve it from a novel perspective by introducing an extra HR contrast image with the same structural information, which provides more effective results. As shown in~\figref{figure1} (a), we can employ the HR T1WI as an auxiliary to improve the resolution of the LR T2WI. More specifically, our MINet accepts two images as input, \ie, an HR T1WI $\mathbf{x}_{T_1}$ and an LR T2WI $\mathbf{y}_{T_2}$. Each image is processed by two independent convolutional branches to obtain multi-stage feature representations. To take advantage of the complementary information between the two images of different contrast, we fuse the representations at the same stage in the two branches so that the anatomical structure features of the auxiliary branch can be fully propagated to the target SR branch. Moreover, we propose a multi-stage integration component to explore the dependencies of multi-contrast features in different stages and allocate appropriate attention weights to them, yielding a more powerful holistic feature representation for SR.

\subsubsection{Multi-Stage Feature Representation.} In order to obtain the anatomical structure of the HR T1WI, we apply a $3\times3$ convolutional layer to obtain an initial representation $\mathbf{F}_{T_1}^0$. Similarly, the LR T2WI is also processed by a convolutional layer, followed by a sub-pixel convolution to restore the scale of the representation (\ie, $\mathbf{F}_{T_2}^0$).

Subsequently, multi-stage feature representations are learned through a cascade of residual groups~\cite{zhang2018image}. In particular, at the $l^{th}$ stage, we obtain the intermediate features as follows:
\begin{align}
    \mathbf{F}_{T_{1}}^{l}&=\mathcal{F}_{T_1}^l(\mathbf{F}_{T_{1}}^{l-1}),\\
    \mathbf{F}_{T_{2}}^{l}&=\mathcal{F}_{T_2}^l([\mathbf{F}_{T_{1}}^{l-1},\mathbf{F}_{T_{2}}^{l-1}]).\label{eq:2}
\end{align}
Here, $\mathcal{F}_{T_1}^l$ and $\mathcal{F}_{T_2}^l$ represent residual groups at the $l^{th}$ stage of each respective branch, and $[,]$ denotes the concatenation operation. Through Eq.~\eqref{eq:2}, our model leverages T1WI features to guide the learning of the T2WI at each stage, enabling progressive feature fusion. Different from most existing methods~\cite{lyu2020multi,lim2017enhanced}, which only use the representations from the final stage for SR, our model stores all the intermediate features at multiple stages: 
\begin{equation}\label{eq:3}
    \mathbf{F} = [\mathbf{F}_{T_1}^1, \mathbf{F}_{T_2}^1, \mathbf{F}_{T_1}^2, \mathbf{F}_{T_2}^2, \cdots \mathbf{F}_{T_1}^L, \mathbf{F}_{T_2}^L].
\end{equation}
Here, $\mathbf{F}$ denotes a multi-stage feature representation obtained by concatenating all the intermediate features, and $L$ indicates the number of residual groups.

\subsubsection{Multi-Contrast Feature Enhancement.}

While Eq.~\eqref{eq:3} aggregates multi-stage features to achieve a more comprehensive representation, it treats each feature independently and does not fully explore the relations among them. To this end, we propose a multi-stage integration module $\mathcal{F}_{\text{Int}}$ (\S\ref{sec:ml}) which learns the relations among features and uses them to modulate each feature. Formally, we have: $\mathbf{H} = \mathcal{F}_{\text{Int}}(\mathbf{F})$, where $\mathbf{H}$ represents the enriched feature representation. 

In addition, we employ a channel-spatial attention module $\mathcal{F}_{\text{Att}}$ to modulate the feature $\mathbf{F}_{T_2}^L$ as $\mathbf{G}_{T_{2}}=\mathcal{F}_{\text{Att}}(\mathbf{F}_{T_{2}}^L)$. Inspired by the spatial~\cite{woo2018cbam} and channel attention mechanisms~\cite{zhang2018image}, we design the channel-spatial attention module, which reveals responses from all dimensions of the feature maps. As shown in~\figref{figure1} (b), the features output by the residual groups $\mathbf{F}_{T_{2}}^L$ are sent into a 3D convolutional layer to generate attention map $\mathbf{A}_{tt}$ by capturing joint channels and spatial features. This can be written as
\begin{equation}\label{eq:10}
\mathbf{G}_{T_{2}}= \texttt{sigmoid}\left(\mathbf A_{tt}\right) \odot \mathbf{F}_{T_{2}}^L+\mathbf{F}_{T_{2}}^L,
\end{equation}
where $\texttt{sigmoid}$ is the sigmoid operation, and $\odot$ denotes element-wise multiplication. As a result, Eq.~\eqref{eq:10} provides a feasible way to enrich $\mathbf{F}_{T_2}^L$ by making use of the context within the feature. We also use a residual layer to effectively preserve information from the original feature map. Note that $\mathbf{G}_{T_{1}}$ is also calculated using Eq.~\eqref{eq:10}.

\subsubsection{Image Reconstruction.}
Since our framework is a pre-upsampling SR, we use a convolutional layer $\mathcal{F}_{\text{Rec}}$ to obtain the final restored image $\hat{{\mathbf x}}_{T_2}$, which can be written as:
\begin{equation}\label{eq:6}
\begin{aligned}
\hat{{\mathbf x}}_{T_2}&=\mathcal{F}_{\text{Rec}}(\mathbf{F}_{T_{2}}^{0}\oplus\mathbf{G}_{T_{2}}\oplus\mathbf{H}), 
\end{aligned}
\end{equation}
where $\oplus$ denotes element-wise summation. In addition, as shown in~\figref{figure1}, we also compute the reconstruction of the T1WI using a convolutional layer with input $\mathbf{G}_{T_{1}}$. Note that the reconstruction procedure is only used as an auxiliary.

\subsubsection{Loss Function.}
We simply use the $L_1$ loss to evaluate the reconstruction results of the T1WI and T2WI. The final loss function is
\begin{equation}
L=\frac{1}{N} \sum_{n=1}^{N} \alpha\left\|\hat{{\mathbf x}}_{T_2}^{n}-{{\mathbf x}}_{T_2}^{n}\right\|_{1}+\beta\left\|\hat{{\mathbf x}}_{T_1}^{n}-{{\mathbf x}}_{T_1}^{n}\right\|_{1},
\end{equation}
where $\alpha$ and $\beta$ weight the trade-off between the T1WI and T2WI reconstruction.

\begin{figure}[!t]
\centering
  \includegraphics[width=0.99\textwidth]{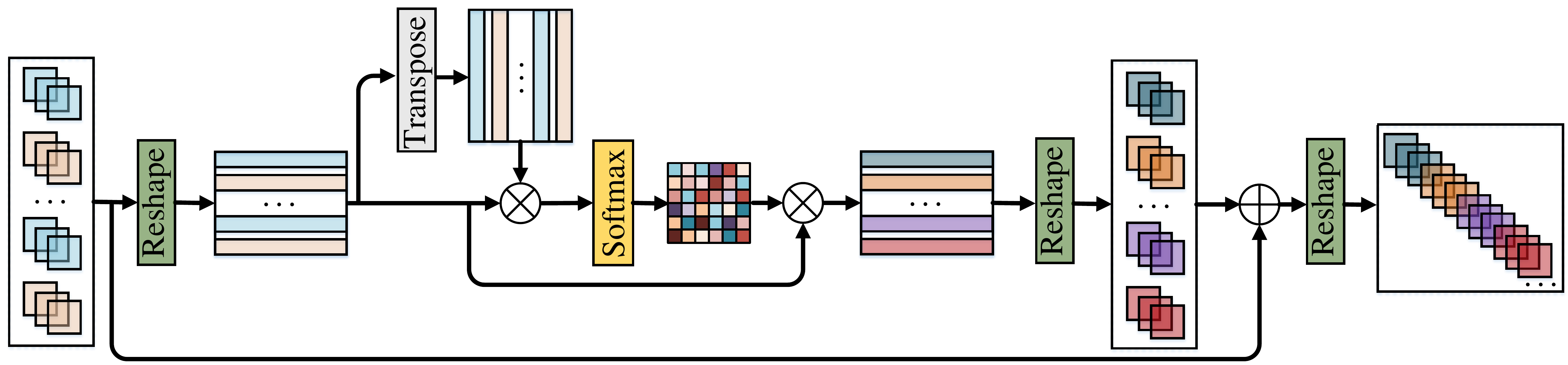}
  \put(-344,70){\scriptsize $\mathbf{F}\!\in\!\mathbb{R}^{2L\!\times\!H\times\!W\!\times\!C}$}
  \put(-303,12){\scriptsize $\hat{\mathbf{F}}\!\in\!\mathbb{R}^{2L\!\times\! HWC}$}
  \put(-247,80){\scriptsize $\hat{\mathbf{F}}^\top\!\in\!\mathbb{R}^{2L\!\times\! HWC}$}
  \put(-200,47){\scriptsize $\mathbf{S}\!\in\!\mathbb{R}^{2L\!\times\! 2L}$}
  \put(-165,12){\scriptsize $\mathbf{S}\hat{\mathbf{F}}\!\in\!\mathbb{R}^{2L\!\times\!HWC}$}
  \put(-50,58){\scriptsize $\mathbf{H}\!\in\!\mathbb{R}^{H\!\times W\!\times\!2LC}$}
  \put(-116,70){\scriptsize $\mathbf{R}{\mathbf{F}}\!\in\!\mathbb{R}^{2L\times \!H\times \!W\!\times\!C}$}   
  \caption{Architecture of the proposed multi-stage integration module.} 
  \label{figure2} 
\end{figure}
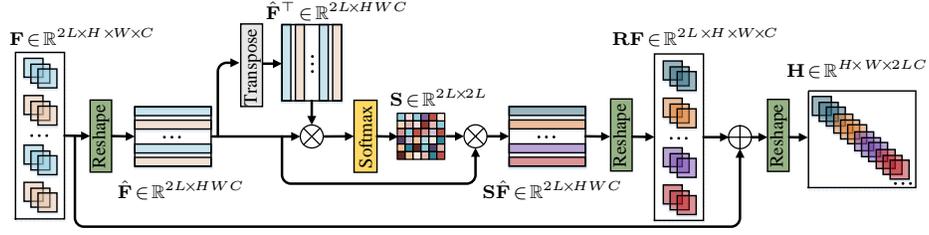
\subsection{Multi-Stage Integration Module} \label{sec:ml}

Although we have aggregated the T1WI and T2WI features in the intermediate stages for T2WI SR, the correlation among different stages has been missed, especially for MRI in different contrasts. In multi-contrast SR, different contrasts can be considered as different classes, and features from different stages can be considered as responses to specific classes. Therefore, we design a multi-stage integration module. As shown in~\figref{figure2}, by obtaining the dependencies between features of the T1WI in each stage with other stages as well as the T2WI in each stage, the network can allocate different attention weights to the features of different contrasts and stages. Thus, while the network pays attention to the more informative stages in the T2WI, it also emphasizes the stages in the T1WI that can guide the SR of the T2WI. 

Recall that $\mathbf{F}\!\in\mathbb{R}^{2L\times\!H\times\!W\times\!C}$ (Eq.~\eqref{eq:3}) concatenates all intermediate representations together. The multi-stage integration module aims to learn the correlations between each pair of features. In particular, $\mathbf{F}$ is first flattened into a matrix representation $\hat{\mathbf{F}}\in\mathbb{R}^{2L\times HWC}$ for computational convenience. Then, we find the correspondence between each pair of features $\hat{\mathbf{F}}_i\in\mathbb{R}^{HWC}$ and $\hat{\mathbf{F}}_j\in\mathbb{R}^{HWC}$ using the following bilinear model:
\begin{equation}
\begin{aligned}
\mathbf{S}&=\texttt{softmax}(\hat{\mathbf{F}}  \hat{\mathbf{F}}^\top) \\
&= \texttt{softmax}([\hat{\mathbf{{F}}}_1, \cdots, \hat{\mathbf{F}}_{2L}][\hat{\mathbf{F}}_1, \cdots, \hat{\mathbf{F}}_{2L}]^\top) \in [0,1]^{2L\times 2L}.
\end{aligned}
\end{equation}
Here, $\mathbf{S}$ denotes the affinity matrix which stores similarity scores corresponding to each pair of features in $\hat{\mathbf{F}}$, \ie, the $(i,j)^{th}$ element of $\hat{\mathbf{F}}$ gives the similarity between $\hat{\mathbf{F}}_i$ and $\hat{\mathbf{F}}_j$. $\texttt{softmax}(\cdot)$ normalizes each row of the input. Next, attention summaries are computed as $\mathbf{S}\hat{\mathbf{F}}\in\mathbb{R}^{2L\times HWC}$ and used to generate an enhanced representation in the following residual form:
\begin{equation}
    \mathbf{H} = \texttt{reshape}(\mathbf{R}{\mathbf{F}} + \mathbf{F}) \in \mathbb{R}^{H\times W\times 2LC},
\end{equation}
where $\mathbf{R}{\mathbf{F}}$ is reshaped by $\mathbf{S}\hat{\mathbf{F}}$. 
In this way, our model is able to learn a more comprehensive and holistic feature representation $\mathbf{H}$ by exploring the relations between features of multi-contrast images at multiple stages.

\section{Experiments} \label{sec:result}
\subsubsection{Datasets and Baselines.}
We evaluate our approach on three datasets: 1) \textbf{fastMRI}~\cite{zbontar2018fastmri} is the largest open-access MRI dataset. Following~\cite{xuan2020learning}, we filter out 227 and 24 pairs of PD and FS-PDWI volumes for training and validation. 2) \textbf{SMS} was constructed using a 3T Siemens Magnetom Skyra system on 155 patients. Each MRI scan includes a T1WI and a T2WI with full $k$-space sampling (TR$_{T1}$ = 2001ms, TE$_{T1}$ = 10.72 ms, slice thickness = 5 mm, matrix = 320$\times$320$\times$20, field of view (FOV) = 230$\times$200 mm$^2$, TR$_{T2}$ = 4511 ms, TE$_{T2}$ = 112.86ms). 3) \textbf{uMR} was constructed using a 3T whole body scanner (United Imaging Healthcare uMR 790\footnote{Provided by {\color{red}{{\href{United Imaging Healthcare, Shanghai, China}{United Imaging Healthcare, Shanghai, China}.}}}}) on 50 patients. Each MRI scan includes a T1WI and a T2WI with full $k$-space sampling (TR$_{T1}$ = 2001 ms, TE$_{T1}$ = 10.72 ms, slice thickness = 4 mm, matrix = 320$\times$320$\times$19, FOV = 220$\times$250 mm$^2$, TR$_{T1}$ = 4511 ms, TE$_{T1}$ = 112.86 ms).
The collection of the clinical datasets was approved by the Institutional Review Board. We split \textbf{SMS} and \textbf{uMR} patient-wise into training/validation/test sets with a ratio of 7:1:2. For \textbf{fastMRI}, we ues the PDWI to guide the SR of FS-PDWI, while for all other datasets we use the T1WI to guide the T2WI. We compare our method with two single-contrast SR methods (EDSR~\cite{lim2017enhanced}, SMORE~\cite{zhao2018deep}) and two multi-contrast methods (Zeng \textit{et al.}~\cite{lyu2020multi}, Lyu \textit{et al.}~\cite{zeng2018simultaneous}).

\subsubsection{Experimental Setup.}
We implement our model in PyTorch with two NVIDIA Tesla V100 GPUs and 32GB of memory per card. Our model is trained using the Adam optimizer with a learning rate of 1e-5, for 50 epochs. The parameters $\alpha$ and $\beta$ are empirically set to 0.3 and 0.7, respectively. We use $L$ = 6 residual groups in out network. All the compared methods are retrained using their default parameter settings.

\subsubsection{Quantitative Results.}
Table~\ref{t1} reports the average SSIM and PSNR scores with respect to different datasets under 2$\times$ and 4$\times$ enlargement. As can be seen, our approach yields the best results on all datasets. This demonstrates that our model can effectively fuse the two contrasts, which is beneficial to the restoration of the target contrast. Notably, the single-contrast SR methods, \eg~EDSR~\cite{lim2017enhanced} and SMORE~\cite{zhao2018deep}, are far less effective than the multi-contrast models. More importantly, however, the multi-contrast SR models,~\ie, Zeng \textit{et al.}~\cite{lyu2020multi} and Lyu \textit{et al.}~\cite{zeng2018simultaneous}, are also less effective than our model, since they do not mine fused features of different contrast or the interaction between different modes at each stage. In particular, when the scaling factor is 2$\times$ on $\textbf{fastMRI}$, we improve the PSNR from 29.484 dB to 31.769 dB, and SSIM from 0.682 to 0.709, as compared to the current best approach, \ie, Lyu \textit{et al.}~\cite{zeng2018simultaneous}. Although it is more difficult to restore images under 4$\times$ enlargement than 2$\times$, our model still outperforms previous methods in extremely challenging settings, which can be attributed to its strong capability in multi-contrast image restoration. 

\begin{table*}[t]
 \centering
 \caption{Quantitative results on three datasets with different enlargement scales, in terms of SSIM and PSNR. The best and second-best results are marked in red and blue, respectively. }
 \resizebox{\textwidth}{!}{
 \setlength\tabcolsep{1.5pt}
 \renewcommand\arraystretch{1.4}
 \begin{tabular}{r||cc|cc|cc|cc|cc|cc}
 \hline
  {Dataset} & \multicolumn{4}{c|}{{{fastMRI}~\cite{zbontar2018fastmri}} }  &  \multicolumn{4}{c|}{{SMS} } & \multicolumn{4}{c}{{uMR} } \\ \cline{1-13}
  {Scale} & \multicolumn{2}{c|}{2$\times$} & \multicolumn{2}{c|}{4$\times$}  & \multicolumn{2}{c|}{2$\times$} & \multicolumn{2}{c|}{4$\times$} & \multicolumn{2}{c|}{2$\times$} & \multicolumn{2}{c}{4$\times$} \\ \cline{1-13}
   Metrics    & PSNR & SSIM & PSNR & SSIM & PSNR & SSIM & PSNR & SSIM & PSNR & SSIM & PSNR & SSIM \\ \hline\hline
  Bicubic &16.571 &0.459  &13.082 &0.105 &21.548  &0.780   &19.508  &0.706  &21.107  &0.730  &19.072  &0.720   \\
  EDSR~\cite{lim2017enhanced} &26.669 &0.512 &18.363 &0.208 &36.415  &0.962   &31.484  &0.886  &35.394  &0.965  &31.165  &0.907    \\
  SMORE~\cite{zhao2018deep} &28.278 &0.667 &21.813 &0.476 &38.106  &0.972   &32.091  &0.901  &36.547  &0.972  &31.971  &0.918   \\
  Zeng \textit{et al.}~\cite{lyu2020multi} &28.870 &0.670 &23.255 &0.507 &38.164  &0.973   &32.484  &0.912  &36.435  &0.971  &31.859  &0.921   \\ 
  Lyu \textit{et al.}~\cite{zeng2018simultaneous} &{\color{blue}29.484} &{\color{blue}0.682} &{\color{blue}28.219} &{\color{blue}0.574} &{\color{blue}39.194} &{\color{blue}0.978}   &{\color{blue}33.667}  &{\color{blue}0.931}  &{\color{blue}37.139}  &{\color{blue}0.977}  &{\color{blue}32.231}  &{\color{blue}0.929}   \\ \hline
   \textbf{MINet} &{\color{red}31.769} &{\color{red}0.709} &{\color{red}29.819} &{\color{red}0.601} &{\color{red}40.549}  &{\color{red}0.983}   &{\color{red}35.032}  &{\color{red}0.948}  &{\color{red}37.997}  &{\color{red}0.980}  &{\color{red}34.219}  &{\color{red}0.956}   \\ \hline  
 \end{tabular}}
 \label{t1}
\end{table*}

\subsubsection{Qualitative Evaluation.}

\begin{figure}[!t]
\centering
  \includegraphics[width=1\textwidth]{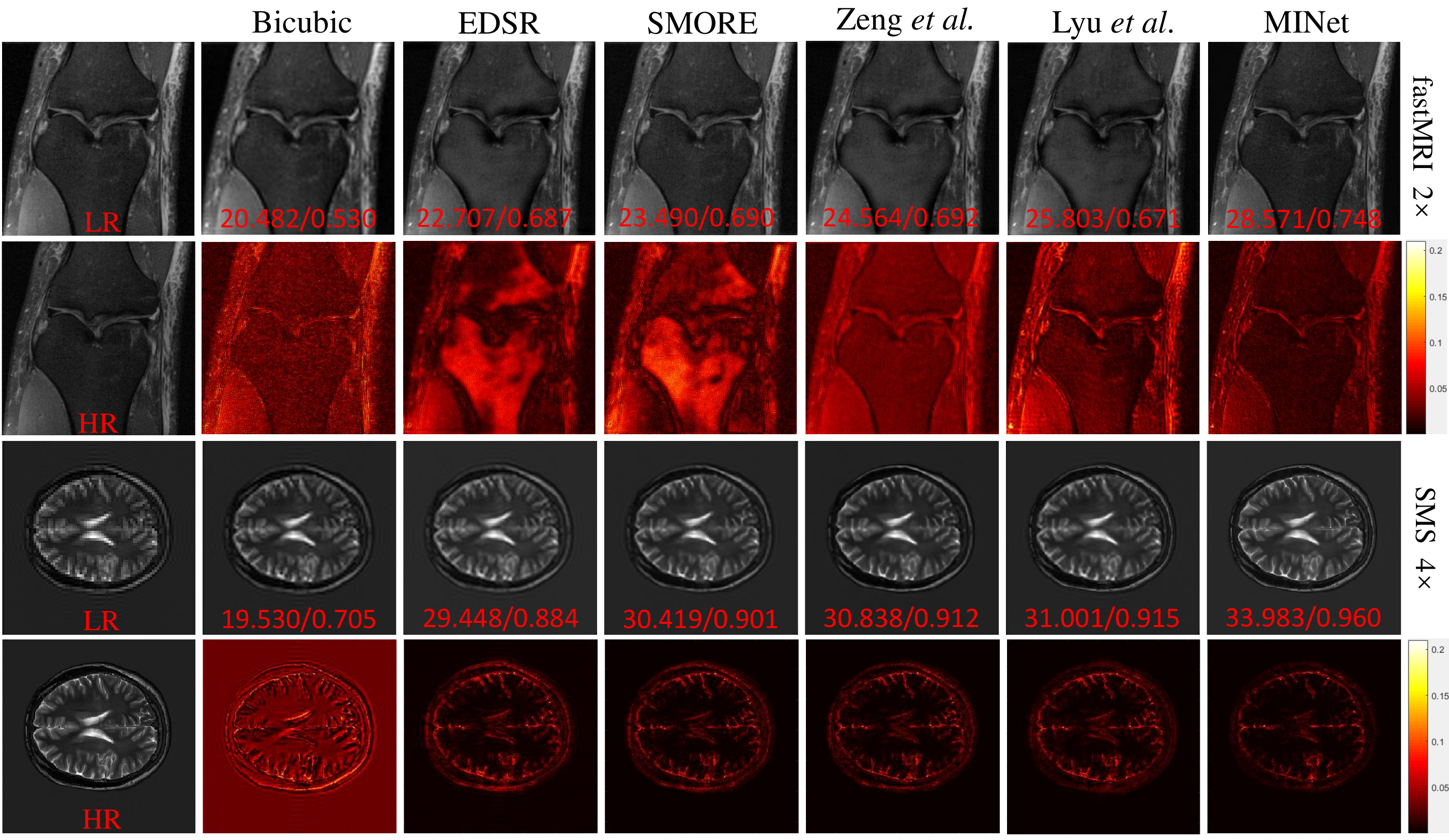}
  \caption{Visual   reconstruction results and error maps of different methods.} 
  \label{figure4} 
\end{figure}

~\figref{figure4} provides the 2$\times$ and 4$\times$ enlargement of target-contrast images and their corresponding error maps on the $\textbf{fastMRI}$ and $\textbf{SMS}$ datasets, respectively. The more obvious the texture in the error map, the worse the restoration. As can be seen, the multi-contrast methods outperform the single-contrast methods. However, our restoration exhibits less of a chessboard effect and fewer structural losses compared to other methods, which is owed to the fact that our model can effectively learn aggregated features from multiple stages. More importantly, the error maps for different enlargement scales demonstrate the robustness of our method across various datasets.

\subsubsection{Ablation Study.}

\begin{table*}[t]
 \centering
 \caption{Ablation study on the {SMS} dataset with 2$\times$ enlargement. The best and second-best results are marked in red and blue, respectively}
 \resizebox{\textwidth}{!}{
 \setlength\tabcolsep{16pt}
 \renewcommand\arraystretch{1.2}
				\begin{tabular}{r||ccc}
			        \hline
			        \hline

			        {Variant} &~~NMSE$\downarrow$~~ &~~PSNR$\uparrow$~~ & SSIM$\uparrow$  \\ \hline\hline
                    $\textit{w/o}$ $\mathcal{F}_{\text{Aux}}$ &0.0037$\pm$0.012 &38.5731$\pm$0.015 &0.9700$\pm$0.009 \\ 
                    \textit{w/o} $\mathcal{F}_{\text{Int}}$ &0.0025$\pm$0.002 &39.2001$\pm$0.108 &0.9776$\pm$0.025 \\ 
                    \textit{w/o} $\mathcal{F}_{\text{Att}}$ &{\color{blue}0.0022$\pm$0.236} &{\color{blue}39.2654$\pm$0.001} &{\color{blue}0.9791$\pm$0.037}\\ \hline
                    \textbf{MINet}  &{\color{red}0.0018$\pm$0.004} &{\color{red}40.5491$\pm$0.014} &{\color{red}0.9830$\pm$0.009} \\ \hline
	        	\end{tabular}}
 \label{t2}
\end{table*}
In this section, we evaluate the effectiveness of the key components of our model through an ablation study. We construct three models: $\textit{w/o}$ $\mathcal{F}_{\text{Aux}}$, which is our model without auxiliary contrast, $\textit{w/o}$ $\mathcal{F}_{\text{Int}}$, which is our model without $\mathcal{F}_{\text{Int}}$, and $\textit{w/o}$ $\mathcal{F}_{\text{Att}}$, which is our model without $\mathcal{F}_{\text{Att}}$. We summarize the 2$\times$ enlargement results on $\textbf{SMS}$ in Table~\ref{t2}. From this table, we observe that single-contrast $\mathcal{F}_{\text{Aux}}$ performs the worst, which is consistent with our conclusion that auxiliary contrast can provide supplementary information for the SR of the target contrast. Since $\mathcal{F}_{\text{Int}}$ reflects the interaction between two contrasts at different stages, the results of $\textit{w/o}$ $\mathcal{F}_{\text{Int}}$ are not optimal. Further, $\textit{w/o}$ $\mathcal{F}_{\text{Att}}$ outperforms $\textit{w/o}$ $\mathcal{F}_{\text{Int}}$ because $\mathcal{F}_{\text{Att}}$ only enhances the features of a single-contrast, and cannot learn the interaction between two different contrasts. Finally, our full MINet, which enhances both the single-contrast feature from $\mathcal{F}_{\text{Att}}$ and the fused feature from $\mathcal{F}_{\text{Int}}$, yields the best results, demonstrating its powerful capability in mining crucial information to guide the target-contrast restoration.

\section{Conclusion}
We have proposed a multi-stage integration network for multi-contrast MRI super-resolution, with the aim of effectively restoring the target-contrast image with guidance from a complementary contrast. Specifically, our model explores modality-specific properties within each modality, as well as the interaction between multi-contrast images. $\mathcal{F}_{\text{Int}}$ and $\mathcal{F}_{\text{Att}}$ are collaboratively applied to multi-level and multi-contrast features, helping to capture more informative features for the target-contrast image restoration. This work provides a potential guide for further research into the interaction between multi-contrast images for MRI super-resolution.

%
%
%
\bibliographystyle{splncs04}
\bibliography{bibliography}

\end{document}